\documentstyle[12pt]{article}
\input{psfig}
\begin{document}
\begin{flushright}
IITAP-96-12-02 \\
December 1996  \\
Revised: September 1997 \\
hep-ph/9709433
\end{flushright}
\vspace*{1.5cm}
\begin{center}
{\large \bf A BFKL Pomeron Manifestation in Inclusive Single Jet 
Production at High-Energy Hadron Collisions} \\
\vspace{1cm}
{\large  Victor T. Kim${}^{\S \dagger}$
and Grigorii B. Pivovarov${}^{\S \ddagger}$ }\\
\vspace*{0.5cm}
{\em $\S $ : International Institute of Theoretical and Applied Physics,\\
     Iowa State University, Ames, Iowa 50011-3022, USA }\\
\vspace*{0.3cm}
{\em ${}^\dagger$ :
St.Petersburg Nuclear Physics Institute,
 188350 Gatchina,
Russia}\footnote{permanent address; e-mail: $kim@pnpi.spb.ru$}\\
\vspace*{0.3cm}
{\em ${}^\ddagger$ :
Institute for Nuclear
Research, 117312 Moscow, Russia}
\footnote{ permanent address; e-mail: $gbpivo@ms2.inr.ac.ru$}
\end{center}

\vspace*{0.5cm}
\begin{center}
{\large \bf
Abstract}
\end{center}

A discrepancy between new data on inclusive single jet
production at the Fermilab Tevatron and perturbative QCD
is discussed. It is shown that the discrepancy may be accounted 
for by the BFKL Pomeron.

\vspace*{1.5cm}
PACS number(s): 13.87.Ce,13.85.-t

\vspace*{2.5cm}
Physical Revew {\bf D57} (1998) R1341-44 

\newpage

The description of the inclusive production of hadron
jets is one of the successes of pertubative QCD (pQCD).
Quantitative agreement between data and theory has been achieved
for jets produced over a wide kinematical range.
In particular, data from CDF and D$\emptyset$ Collaborations
at the Fermilab Tevatron on inclusive single jet production
for $\sqrt{s}=1800$ GeV  \cite{CDF96a,D096b}
are in agreement with pQCD for jet
transverse energy ranging approximately from 15 to 400 GeV.
Over this transverse energy 
range the cross section falls by seven orders of magnitude.

With this good quantitative description of the dependence
on the parameters of the produced jets at fixed
total energy of the collision, the natural next task is to
examine the dependence of the production cross section on the
total energy.
Dimensional analysis and scaling hypothesis predetermines
this energy dependence. pQCD (see, e.g., review \cite{CTEQ}) 
dictates a particular mechanism of 
scaling violation involving
a hadronic scale, $\Lambda_{QCD}$, which yields a specific non-trivial
 energy dependence. Any deviation from this prediction
would manifest an inadequacy of the pQCD framework for managing
nonperturbative physics (soft hadronic radiation).

We will show that existing data already contain evidence for 
additional non-pQCD effects which are consistent with the
Balitsky-Fadin-Kuraev-Lipatov (BFKL) Pomeron \cite{Lip76} 
framework. We can begin with a more general observation that 
a potential mechanism for such non-pQCD scaling violation
is implied by  resummation of the leading energy logarithms 
of QCD (for a recent review, see Ref. \cite{Lip97}). 
In this case, the dependence of the cross section on the QCD running
coupling constant  
 $\alpha_{S}$ differs from the simple power dependence of pQCD.
This change away from a power dependence
 yields an altered energy
dependence of the cross section.

There are preliminary data from CDF \cite{CDF96b}
on the cross section for inclusive single
jet production  at $\sqrt{s}=1800$ GeV and $\sqrt{s}=630$ GeV.
In particular, we examine the ratio of cross sections 
scaled by the jet transverse energy
$E_{\perp}$, taken at the same values of 
$x_{\perp}=2E_{\perp}/\sqrt{s}$ and the rapidity of the jet $\eta$,
but at different total energies \cite{UA285,CDF93}:
\begin{equation}
 R(x_{\perp},\eta)=\frac 
{\Big( E_{\perp}^{4} E d^{3} \sigma /d^{3} k\Big)
\mid _{\sqrt{s_{1}},x_{\perp},\eta}}
{\Big( E_{\perp}^{4} E d^{3} \sigma /d^{3} k\Big)
\mid _{\sqrt{s_{2}},x_{\perp},\eta}}.
\end{equation} 
Without scaling violation the scaled ratio is unity, regardless
of the dynamics.
On the pQCD leading order predictions for the ratio see \cite{KPV97}.

Comparison of the next-to-leading order (NLO) pQCD prediction 
\cite{Ell90} with the data \cite{CDF96b}
shows a noticeable discrepancy at small $x_{\perp}$ (Fig. 1). 
This problem has been already seen in previous CDF data 
\cite{CDF93}. It is unlikely that complications connected 
with jet algorithms and various uncertainties of pQCD approach  
\cite{Ell93} may account for the discrepancy.

\begin{figure}[htb]
\vskip 9 cm
\includegraphics{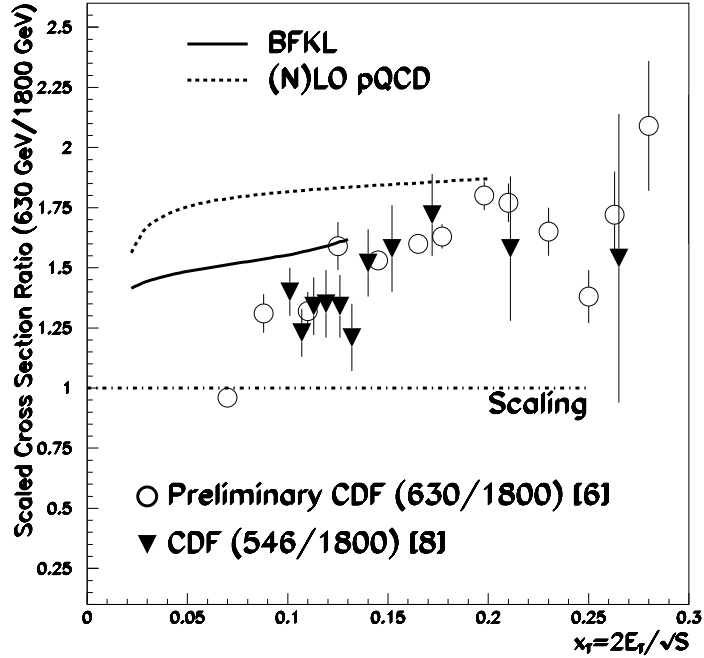}
\caption{The $x_{\perp}$-dependence of the scaled 
cross section ratio (630 GeV)/(1800 GeV). 
Only statistical errors are shown.}
\end{figure}

In this paper we show that the discrepancy in the scaled ratio
at small $x_{\perp}$ is accounted
for by the BFKL Pomeron \cite{Lip76}.

In our approach, the dependence of the inclusive jet 
cross section on $\alpha_{S}$ may be expressed
 as a multiple integration and summation 
over conformal dimensions and conformal spins 
of the BFKL Pomerons \cite{KP96b}.
Each term of this "sum" depends on $\alpha_{S}$ as 
$x_{\perp}^{-\alpha_{S} \beta(\{\nu,n\}) }$, where
$\{ \nu, n\} $ is a set of  conformal dimensions, $\nu_{i}$,
 and conformal spins, $n_{i}$. To get the weights with which 
these contributions enter the "sum" as well as the $\beta(\{\nu,n\})$,
we use the effective Feynman-like rules defined in Ref. \cite{KP96b}.
Substituting  the running $\alpha_{S}$
into the BFKL formulas for the inclusive cross sections
one obtains a new $x_{\perp}$-dependence for the 
scaled ratio $R$.
The calculation of the scaled ratio 
along this line gives the result presented in Fig. 1.

 As was pointed out in Ref. \cite{Mue87}, it is
important to keep track of the most forward and backward
jets of the events for the BFKL kinematics. In particular 
\cite{KP96b}, the inclusive single jet cross section is a sum
of three terms. The first (second) term comes from the processes 
with untagged most forward (most backward) jet and corresponds to the
diagram of Fig. 2a (Fig. 2b). The third term (Fig. 2c) corresponds 
to the processes with both most forward and most backward jets
untagged. The analytic expressions corresponding to the diagrams
of Fig. 2 read as follows:
\begin{figure}[htb]
\vskip 3.7 cm
\includegraphics{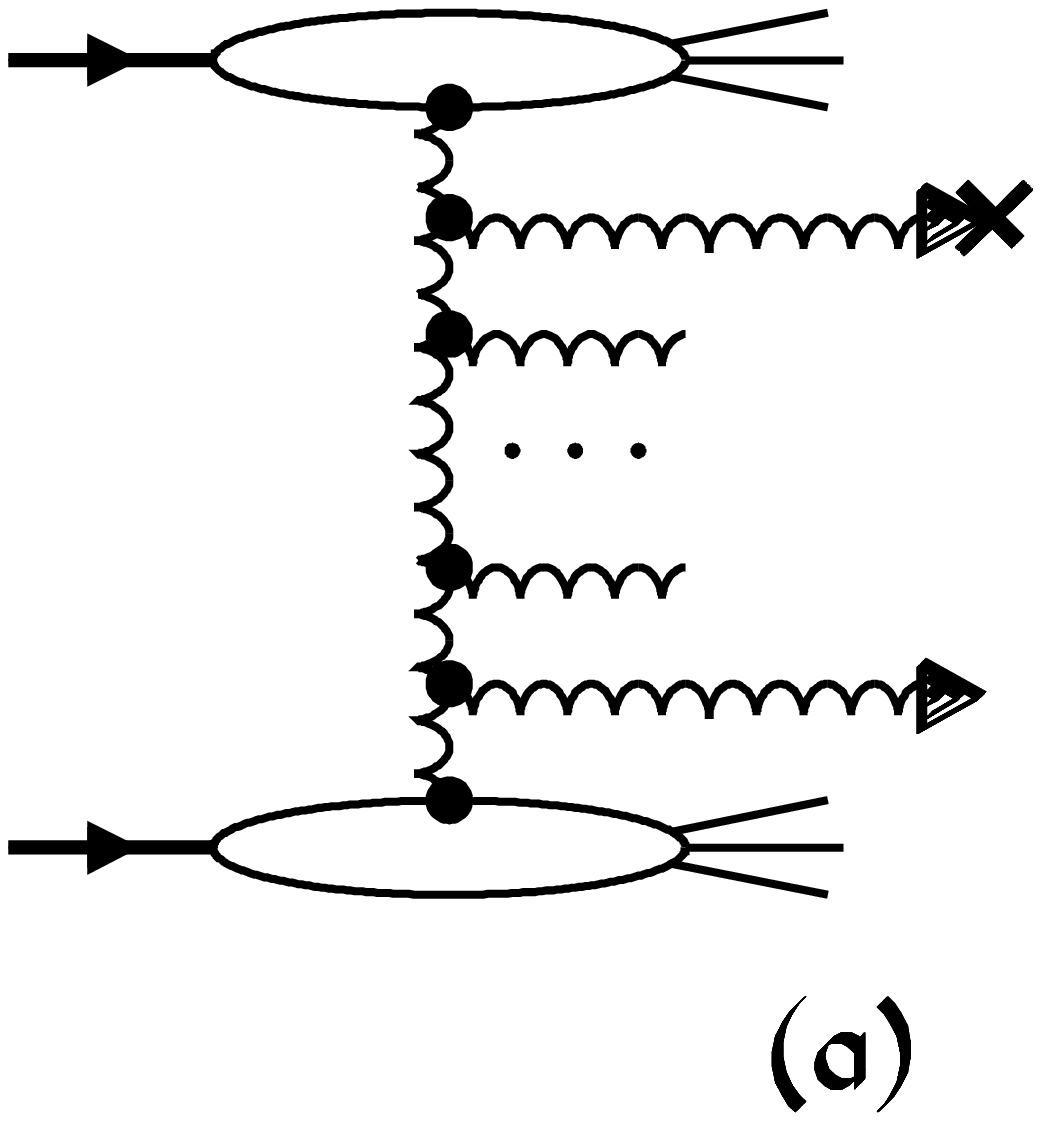}
\includegraphics{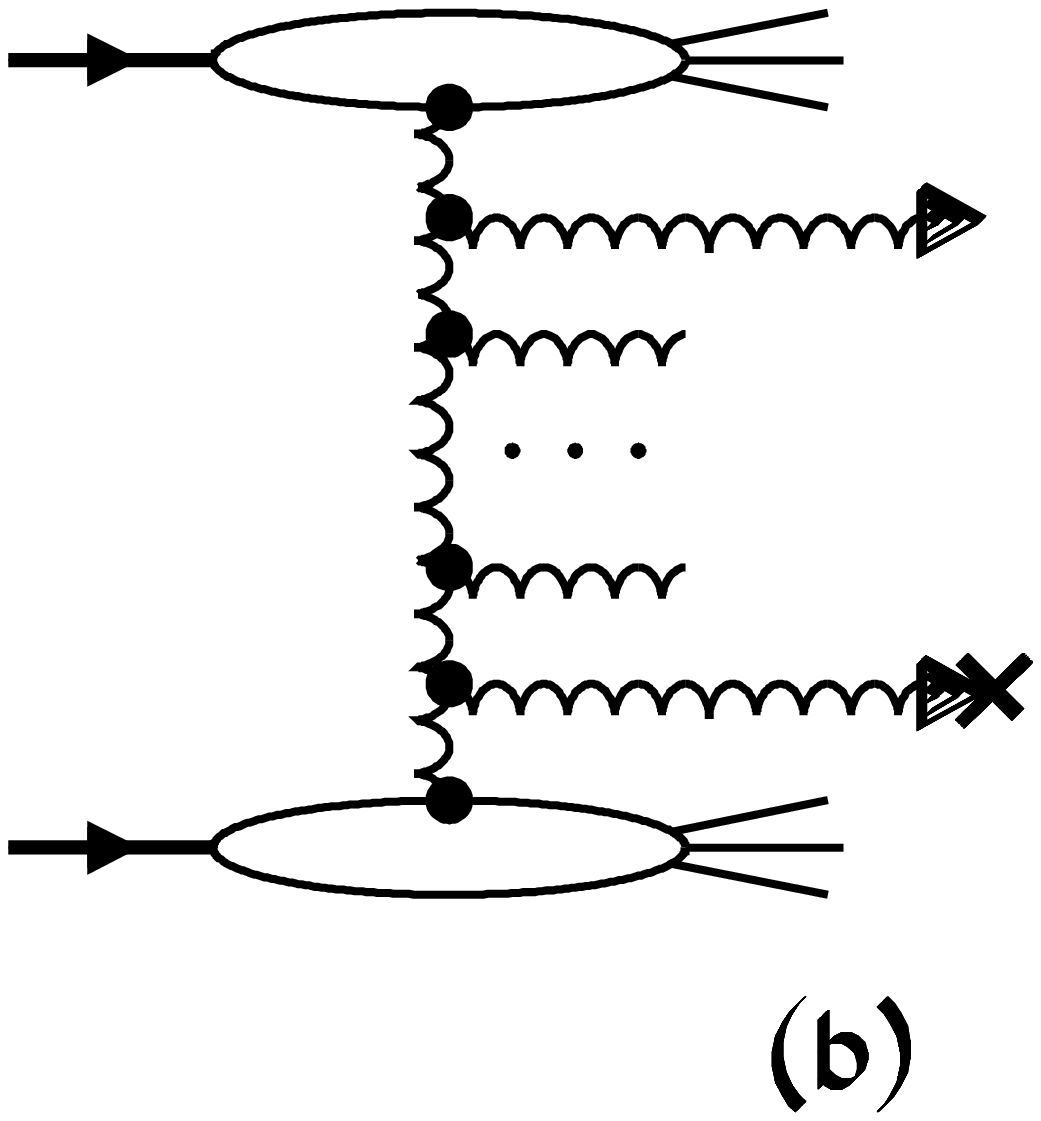}
\vskip 9.2 cm
\includegraphics{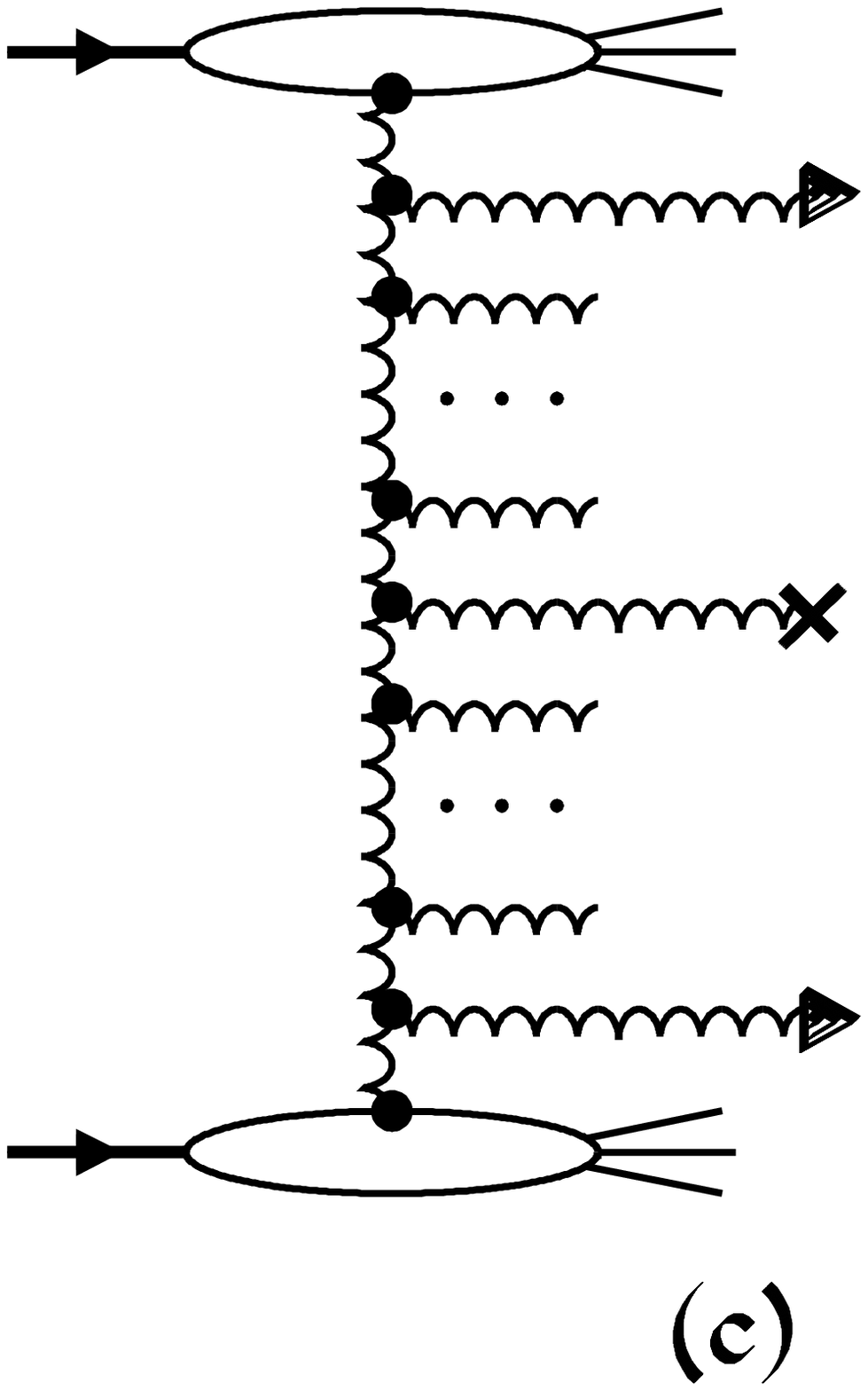}
\vskip 1cm
\caption{Diagrams for inclusive single jet production.
$\times$ denotes
tagged jet, $\rhd$ marks most forward(backward) jet.}
\end{figure}
\begin{equation}
\frac{\alpha_{S} N_{c}}{2 \pi^{2}}
\, \vartheta (x^{+}-x^{-}) \, F_{B} (e^{-(x^{-}-x_{B}^{-})}, \mu_H)
 \int_{- \infty}^{\infty} \! \! \! d \nu 
\, \, \tilde{W}_{A} (x_{A}^{+}-x^{+},x_{A}^{+}-x^{-}, \nu, \mu_H ),
\end{equation}

\begin{equation}
\frac{\alpha_{S} N_{c}}{2 \pi^{2}}
\, \vartheta (x^{+}-x^{-}) \, F_{A} (e^{-(x^{+}_{A} - x^{+})}, \mu_H)
 \int_{- \infty}^{\infty} \! \! \! d \nu 
\, \, \tilde{W}_{B} (x^{+}-x^{-}_{B},x^{+}-x^{-}_{B}, \nu , \mu_H),
\end{equation}

\begin{eqnarray}
&& \frac{\alpha_{S} N_{c}}{2 \pi^{2}} \, 
 \int \! \! \! \int_{- \infty}^{\infty} \! \! \!
 d \nu_{1} \, d \nu_{2} 
\, \, W_{A} (x_{A}^{+}-x^{+},x_{A}^{+}-x^{-}, \nu_{1}, \mu_H)
\nonumber \\
& & \times \,  R_{\varphi}(0, -\nu_{1}-\nu_{2})
 \, W_{B} (x^{+}-x^{-}_{B},x^{+}-x^{-}_{B}, \nu_{2} , \mu_H),
\end{eqnarray}
where $x^{\pm}$ are connected with the light-cone components
of the produced jet momentum:
 $x^{\pm} = \pm \log \big( {k^{\pm}}/{\mu_R} \big) $;
 $x^{\pm}_{A,B}$ are connected with the light-cone components
of the momenta for the colliding hadrons $A$ and $B$:
 $x^{\pm}_{A,B} = \pm \log \big( {k^{\pm}_{A,B}}/{\mu_R} \big) $;
$F_{A,B}$ are the effective parton densities of the colliding hadrons
$A$ and $B$; $\mu_H$ is the normalization point for both 
the parton densities
and the running coupling constant $\alpha_{S}$ (subscript $H$ stands
for ``Hard''); $\mu_R$ is the normalization point for the energy 
logarithms which are resummed by the BFKL Pomeron (subscript $R$ 
stands for ``Regge'').
Integrations over conformal dimentions $\nu_i$ \cite{Lip76} are to 
account for different patterns of gluon radiation occupying the rapidity
intervals spanned by the tagged jet and the most forward (backward)
jet. Note that the radiation involves infinite number of radiated
gluons.
The following analytic expressions for $W_{A,B}$, $\tilde{W}_{A,B}$
may be obtained from the diagram technique of Ref. \cite{KP96b} 
by integration over parameters of the untagged most forward or(and) 
untagged most backward jet(s):

\begin{eqnarray}
W_{A,B}(x_{1},x_{2},\nu,\mu_H)=\tilde{W}_{A,B}(x_{1},x_{2},\nu,\mu_H)
 \frac {\Gamma (\frac {1}{2} - i \nu)}{\Gamma (\frac{1}{2}+ i \nu)},
\end{eqnarray}
\begin{eqnarray}
\tilde{W}_{A,B} (x_{1},x_{2}, \nu, \mu_H) & = & 
\frac{\alpha_{S} N_{c}}{2 \pi^{2}} \,
i \frac {e^{ \big(- i x_{1} (-\nu + i \frac {1+\omega(0,\nu )}{2}) \big)}}
  { \nu + i \frac {1+\omega(0,\nu )}{2} } \nonumber \\
& \times &\bigg[ e^{ \big(- i x_{2} 
(- \nu + i \frac {1+\omega(0,\nu )}{2})  \big)}
M_{A,B}(1+\omega(0,\nu),\mu_H,m(x_{1},x_{2})) \nonumber \\ 
& - & M_{A,B}(\frac{1 + \omega(0,\nu)}{2} + i \nu, \mu_H,m(x_{1},x_{2}))
 \bigg],
\end{eqnarray}
where $\omega(0,\nu)$ is the Lipatov eigenvalue  
$\omega(n,\nu) = ({2 \alpha_{S} N_c}/{\pi})
\Big( \psi(1) - Re \, \psi \big( \frac{|n|+1}{2} +
i\nu \big) \Big)$ \cite{Lip76} taken
at zero value of the conformal spin $n$ ($\psi$ here 
is the logarithmic derivative of the Euler Gamma-function);
\begin{equation}
M_{A,B} (\lambda, \mu_H, \gamma) = \int_{e^{-\gamma}}^{1}
 \! \! \! d z \, z^{\lambda-1} \, F_{A,B}(z,\mu_H)  
\end{equation}
are incomplete moments of the parton densities; and
$ m(x_{1},x_{2})=min \{ x_{1},x_{2} \} $.

One more object, $R_{\varphi} (0, \nu)$, entering Eq. 4
is an element of the diagram technique of \cite{KP96b},
\begin{equation}
R_{\varphi} (n, \nu) = 
\frac{i^{|n|} e^{i |n| \varphi}}{\frac{|n|}{2}-i (\nu +i \varepsilon)} \,\,
\frac{\Gamma (\frac{|n|}{2}+1 - i \nu)}{\Gamma (\frac{|n|}{2}+1 + i \nu)}, 
\end{equation}
taken at $n = 0$.

The sum of Eqs. 2-4 gives the cross section
${ (s/\pi^{4}) d^{2} \sigma}/ {dx^{+} dx^{-}}$ which is easily
connected to the invariant cross sections entering Eq. 1.

The first calculational task is to compute the incomplete
momenta of the parton densities entering Eq. 6.
For the effective parton densities, $F=g + 4/9 (q + \bar{q})$
\cite{Com84}, we use the following parameterization
\begin{eqnarray}
F (z,\mu_H)=  a(\mu_H) z^{-b(\mu_H)}
(1-z)^{c(\mu_H)},
\end{eqnarray}
where we omitted the subscripts $A$ and $B$, since,
in our case, both colliding hadrons ($p$ and $\bar{p}$) have
the same effective parton density.
With this parameterization the incomplete moments 
are analytically calculable as 
\begin{eqnarray}
M (\lambda,\mu_H,\gamma) & = &  a
(\mu_H)  \Big( B (\lambda - b(\mu_H),c(\mu_H)+1)
  \nonumber \\ 
& - & \frac{e^{\gamma (b(\mu_H)-\lambda )}}{\lambda -b(\mu_H)}
\Phi ( \lambda-b(\mu_H), -c(\mu_H), \lambda -b
(\mu_H)+1, e^{-\gamma} )
 \Big),
\end{eqnarray}
where B is the Euler Beta-function and $\Phi$ is 
the hypergeometric function. 
Parameters $a(\mu_H),b(\mu_H),c(\mu_H)$  were found by
fitting the CTEQ4L \cite{CTEQ97} parton densities. 

The next step is to perform the integration of Eqs. 2-4 over
conformal dimensions. This has been done numerically.
The relative error of this numerical computation
does not exceed $10 \%$ for the results presented, 
though, for most values of the parameters, it is less than 
few percents. 

The only free parameters which we have is the normalization
points $\mu_H$ and $\mu_R$. They were taken to be proportional to the 
transverse
energy: $\mu_{H,R} = \xi_{H,R} E_{\perp} $, as in pQCD \cite{Ell90}.
Our main result is that with fixed values of $\xi_H = 1 $ and $\xi_R = 0.5 $,
one has a  good description of data \cite{CDF96b} 
simultaneously for jet cross 
section at $\sqrt{s} =  630$ and  $1800$ GeV (Fig. 3), 
and for the scaled ratio (630 GeV)$/$(1800 GeV) (Fig. 1). 
As is seen from Fig. 1, pQCD fails to do this for the scaled ratio
at low $x_{\perp} $. (Note, that CDF data \cite{CDF96b}
are still preliminary and require final analysis.)

\begin{figure}[htb]
\vskip 9 cm
\includegraphics{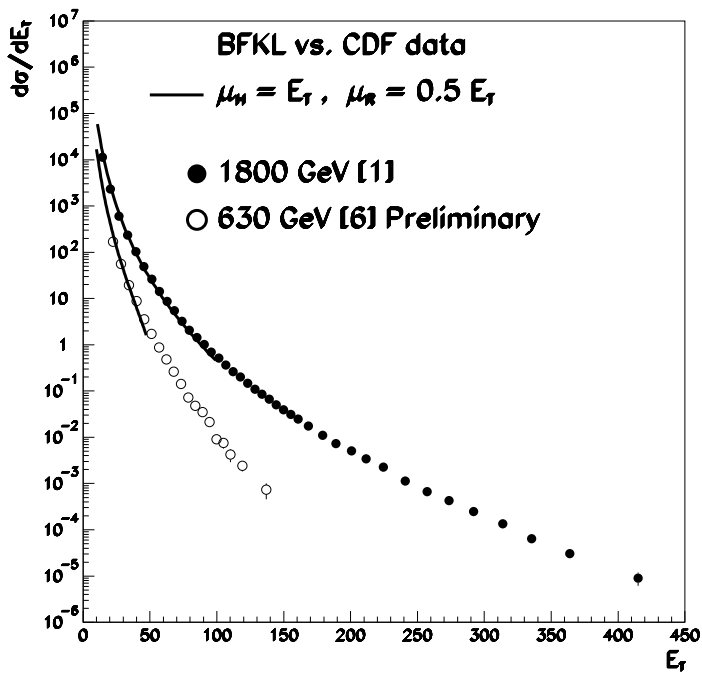}
\caption{Inclusive jet production cross sections for Tevatron energies.}
\end{figure}

Qualitatively, the above result is a consequence of the
energy dependence of the gluon radiation (see Fig. 2)
which is neglected in the finite order pQCD calculations:
The radiation is more significant for higher energies where 
more ``diffusion'' of  the transverse momenta is allowed 
\cite{Lip76,Lip97}.

Here we should comment on the applicabilty of the running
coupling constant in conjunction with the BFKL Pomeron
and  on the normalization point dependence of 
the BFKL predictions.

There are no ultraviolet divergencies
in our resummed leading energy logarithms 
and, hence, there are no
running coupling constant and
ultraviolet renormalization group invariance. 
Presumably, both should arise via resummation of subleading
energy logarithms along with appropriate new factorization 
theorems. Connection between ``Hard'' and ``Regge'' normalization points
should also be established via analysis of subleading energy logarithms.
Our work assumes the eventual success of such a program.
For recent works on resummation of subleading energy logarithms
see Ref. \cite{Fad96} and references therein. In addition, attempts 
to find the relevant versions
of factorization theorems have appeared \cite{Cat91,Bal96,KP96a}.
For more examples of the BFKL Pomeron in phenomenological
applications encountering  the problem of normalization point
dependence see  Refs. \cite{Mue87,KP96a,Rys80}.
We note also that significant normalization point dependence
and a deviation from the data \cite{H195} 
was observed in the calculation of forward jet production
in NLO pQCD \cite{Mir96} for deep inelastic scattering
at the HERA energies, while the BFKL calculation \cite{Bar96}
 agrees with the data \cite{H195}.

Our conclusion is that the BFKL prediction has more capacity
than the prediction of pQCD
in fitting the data on inclusive single jet production
for both $\sqrt{s}=630$ GeV and $\sqrt{s}=1800$ GeV. 
In both approaches, BFKL and pQCD,
there is a fitting parameter $\mu_H$; BFKL approach has additional
parameter, $\mu_R$, wich was also used to fit the data. In the case 
of pQCD the good fit for $\sqrt{s}=1800$ GeV is achieved
at the expense of the fit for the lower energy
at small $x_{\perp}$. However, the BFKL prediction
is able to accommodate both energies.

 We thank A.V. Efremov, L.L. Frankfurt, I.F. Ginzburg, 
 E.A. Kuraev, V.A. Kuzmin, L.N. Lipatov, V.A. Matveev, 
 J. Qiu, J.P. Vary, A.A. Vorobyov, A.R. White, M. W\"usthoff, 
 and D. Zeppenfeld for stimulating discussions.
 We are grateful to the Fermilab Theory Division for warm hospitality.
 V.T.K. is  thankful to S.J. Brodsky, A. Goussiou, J. Huston,
 H.S. Song, N. Varelas and  H. Weerts for fruitful 
 discussions, and to the Aspen Center for Physics for their kind 
 hospitality. G.B.P. wishes to thank L.McLerran and R.Venugopalan for useful 
 discussions, and also the Institute for the Nuclear Theory of
 the  University of Washington, Seattle and the Nordita, Copenhagen 
 for their warm hospitality. 
 We thank the National Energy Research
 Scientific Computing Center (NERSC) at the LBNL for
 high-performance computing resources.
 This work was supported in part by the Russian Foundation
 for Basic Research, grants No. 96-02-16717 and 96-02-18897.

\end{document}